# Topological and morphological signatures of disorder in a self-assembled, soft matter sponge network

Xueyang Feng[1*], Suman S. Kulkarni[2*], Michael S. Dimitriyev[1], Dani S. Bassett[2], Randall D. Kamien[2], Edwin L. Thomas[1†], Gregory M. Grason[3†]

**Affiliations:**
[1]Department of Materials Science and Engineering, Texas A&M University, 209 Reed McDonald Building, College Station, TX 77843, USA
[2]Department of Physics and Astronomy, University of Pennsylvania, 209 S 33rd Street, Philadelphia, PA 19104, USA
[3]Department of Polymer Science and Engineering, University of Massachusetts Amherst, 120 Governors Dr, Amherst, MA 01003, USA
[*]Equal contributors
[†]Corresponding authors. G.M.G (grason@umass.edu) and E.L.T. (elt@tamu.edu)

**Abstract:**
Many soft matter systems exhibit ordered, polycontinuous network morphologies, such as the cubic (double) gyroid or diamond, as well as disordered network morphologies known generically as “random sponges”. While presumed to share similar local packing geometry, the structural relationship between these ordered and disordered network morphologies has remained obscure. We use slice and view scanning electron microscopy to analyze and compare multi-scale morphological features of an ordered double-gyroid morphology to the amorphous sponge morphology formed in the same block copolymer sample. We find that node valence of the minority component network of the sponge is mostly gyroidal (trivalent), with a small fraction of diamond-like (tetravalent) connections. We analyze mesoatoms—space-filling volumes occupied by chains around each network node—finding significant differences in shape and size between ordered and amorphous regions. Local block thickness and inter-domain curvature within mesoatomic units of the disordered sponge exhibits a surprisingly similar degree of dispersity to the ordered double-gyroid. The mean differences in local packing geometry derive from topological distinction: loops of the minority networks of the ordered double-gyroid are intercatenated, while loops of the disordered sponge are not. In this way, the sponge may be viewed as disordered variant of a single-gyroidal morphology. We exploit these topological differences to demarcate the boundary region between ordered and disordered networks and highlight modulations of the mesoatom motifs at the boundary. These observations point to new questions about potential metastability of disordered networks and their possible role as kinetic precursors to long-range ordered network morphologies.

## Introduction

Self-assembled morphologies of molecular amphiphiles constitute a basic category of nanostructured materials from synthetic to biological systems (1, 2). A diverse range of possible amphiphilic molecular structures can form, falling into a few basic categories according to the geometry and topology of local domains into which chemically distinct parts of the molecules segregate. For example, the most intuitive morphologies consist of spherical, cylindrical, or lamellar domains. These basic motifs can in turn arrange themselves into periodic (or nonperiodic) structures at the multi-domain scale. Arguably the most subtle and enigmatic of these domain structures are the polycontinuous network (PCN) morphologies (3, 4), which can be crystalline or amorphous (cPCN or aPCN). In these structures, the chemically segregated subdomains of each type maintain a characteristic local nanoscale thickness and each subdomain is continuously connected throughout the entire structure, unlike spheres, cylinders or layers, which are topologically disconnected. PCN morphology *requires* the unlike chemical domains to be both mutually interperforated *and* delineated by inter-material dividing surfaces (IMDSs) built from saddle-like shapes (1, 5). The integration of distinct chemical nanodomains with continuous topology and especially large IMDS area to volume ratios leads to useful functional properties of materials with PCN structure, with high-value applications spanning from ultrafiltration (6, 7) and batteries (8, 9) to photonic media (10–12).

The best studied forms of PCN possess double-network topology and triply-periodic symmetries (13), and form by equilibrium assembly in amphiphilic liquid crystals and block copolymers. Such cPCNs, like the double-gyroid (DG) or double-diamond (DD), are composed of two tubular network domains (A-type component), meeting in trivalent and or tetravalent nodal junctions, respectively. Separating the tubular networks is a slab-like matrix domain (B-type component), whose saddle-like mid-surface roughly approximates triply-periodic minimal surface (TPMS) shapes, respectively the Schoen's G and Schwarz's D surfaces (14). Linking the structure and arrangements of molecules to this complex mesodomain geometry of cPCNs is made feasible by employing crystallographic symmetry (15). DG and DD can be decomposed into multiple copies of nodal motifs of trihedral or tetrahedral interconnections situated at appropriate Wyckoff positions of the Ia3d and Pn3m space groups, respectively. Notably, even though these nodal motifs, which we denote as the *mesoatoms* of PCN, are uniform in the ideally cubic double-networks, their local geometries (e.g. thicknesses and curvatures) are intrinsically variable, arising from unavoidable conflicts between uniform space-filling and favored domain curvature broadly referred to as *packing frustration* (16–19). Beyond the intrinsic heterogeneity in the ideal cubic structure, detailed analysis of experimental cPCN structures formed in block copolymers show significant departure from perfect cubic symmetry presumably due to long-lived non-equilibrium effects arising due to processing (20). Such highly non-affine deformations from the cubic structure are accommodated by malleability of the mesoatom motifs, resulting in asymmetric distributions of inter-node distances and angles in the networks (21).

In this article, we focus on a comparative study of a related class of PCN morphologies in block copolymers (BCPs) that lack long-range order, for which the connection between molecular-scale (subdomain) and network-scale (supradomain) structure has remained hidden from sight. These aPCNs sometimes referred to as "random sponges", are characterized by locally catenoidal IMDS shapes that separate quasi-tubular and pore-like geometries of the chemically dissimilar components, each of which is continuously connected but seemingly in purely random ways. Such aPCNs appear in a wide variety of amphiphilic systems, from kinetically arrested, non-equilibrium states such as thermal- or solvent-quenched BCPs (22–25), randomly crosslinked (26–28) and phase-separating polymer blends (29–31), and to the equilibrium polymeric microemulsions (32) and L3 membrane phase of lyotropic systems (33–35). Lingering over observations of aPCN in block copolymers, which are known to exhibit long-ranged ordered equilibrium states, is the question of whether the disordered sponge only occurs as a transient precursor to long-range order, or else could it exist as thermodynamically stable, or if not deeply metastable, disordered phase?

Although they lack long-range order, aPCN structures consist of a random, tubular-pore geometry, and hence are characterized by some sort of network structure. That is, just as is the case for ordered cPCNs, the networks threading through tubular domains of one chemical type in aPCNs are characterized by nodes, edges and loops. At present, the structure of these seemingly random networks, and its relationship to the structure of their ordered counterparts, is unknown. Clearly, the lack of long-range order implies that network properties of sponges are statistically distributed, but what type of spatial correlations are encoded in aPCNs, and how do these relate to local geometric features of amphiphilic molecular packing? Prior experimental studies of aPCN morphology have not had direct access to 3D structure across the range of length scales necessary to assess intra-network correlations. This is because most structural studies have relied on either scattering (32, 36–38), whose rather broad spectra of superposed Fourier amplitudes lack direct connection to a unique spatial structure, or else on 2D projection images using electron microscopy of thin sections of the material (39), which while consistent with a random sponge structure do not allow for tracing of the true intra-domain connectivity. A limited number of studies have carried out 3D tomographic reconstructions of aPCN structures, for example by using transmission electron microscope tomography (40–42) . However, such methods are rather limited in terms of the depth of accessible 3D imaging (typically ~100 nm or less) compared to typical nano subdomain size (~ 10s of nm), making it nearly impossible to trace out even a single closed loop of the random tubular network.

Here, we use the considerable advantages afforded by slice-and-view scanning electron microscopy (SVSEM) (20, 5, 43) to characterize the structure of aPCNs formed by a diblock copolymer, quantifying morphological features spanning from subdomain metrics of molecular packing to supradomain features at the network scale. Notably, the disordered sponge formed in this single molecular component system is observed to coexist with an ordered DG phase forming in adjacent sample regions, but from the very same amphiphilic polymer, providing a direct basis for structural comparison between PCNs with and without long-range order. We show that, notwithstanding the lack of long-range order, the aPCN is built from similar mesoatomic nodal elements, predominantly trivalent units randomly interconnected with a small number of tetravalent units. The degree of variability in packing geometry at the molecular scale is found to be surprisingly similar in both the amorphous and ordered PCN structures, with statistically coherent shifts in the average IMDS curvature and subdomain thickness between the two. We show that these morphological features at the subdomain scale can be linked to a topological distinction: in the cPCN, loops of ordered double-network are uniform in size (10 nodes per loop) and are uniformly catenated by loops of another network of the same chemical composition; whereas loops of the aPCN have variable size but are uniformly unlinked. These observations show that aPCN structure of this block copolymer is akin to randomized, single-gyroid morphology, and raises new questions about the thermodynamics of this network-liquid morphology and kinetic pathways from the disordered state to a long-range order PCN structure. Moreover, these results on the multi-scale structure of aPCN in block copolymers point the way to new measures of correlated disorder in random material networks more broadly.

## Results

We analyze and compare self-assembled network morphologies from toluene cast films of polystyrene-*b*-polydimethylsiloxane (PS-PDMS) diblock copolymers. After solvent evaporation, (see **Materials and Methods**) the sample is in a state of partial equilibration, characterized by large grains of well-ordered DG structures (cPCN) separated by narrower, yet substantial, regions of amorphous PCN (aPCN), as revealed by the large field SEM image in **Fig. 1A**. 3D tomographic reconstructions of the morphologies are carried out using the SVSEM technique described previously (20). The higher yield of secondary electrons from the PDMS domains over that from the PS domains provides strong contrast. The voxel resolution in the reconstruction is about 3 nm on a side. To capture intra-network structure at the scale of loops and their catenation, we analyze volumes that are in excess of several hundred nanometers on a side that, as we show below, is several-fold larger than the nanodomains themselves (e.g. the mean diameter of the tubular PDMS network domains is approximately 25 nm, roughly half the caliper dimension of mesoatoms) corresponding to analysis of volumes that include several thousand nodal mesoatoms.

We first consider the direct comparison of two separate regions, both of which are approximately $(450\ \text{nm})^3$, but corresponding respectively to ordered DG and disordered sponge, as shown in 2D SEM sections in **Figs. 1B** and **1E**. We then describe the signatures of the PCN structure at the boundary between ordered and amorphous regions.

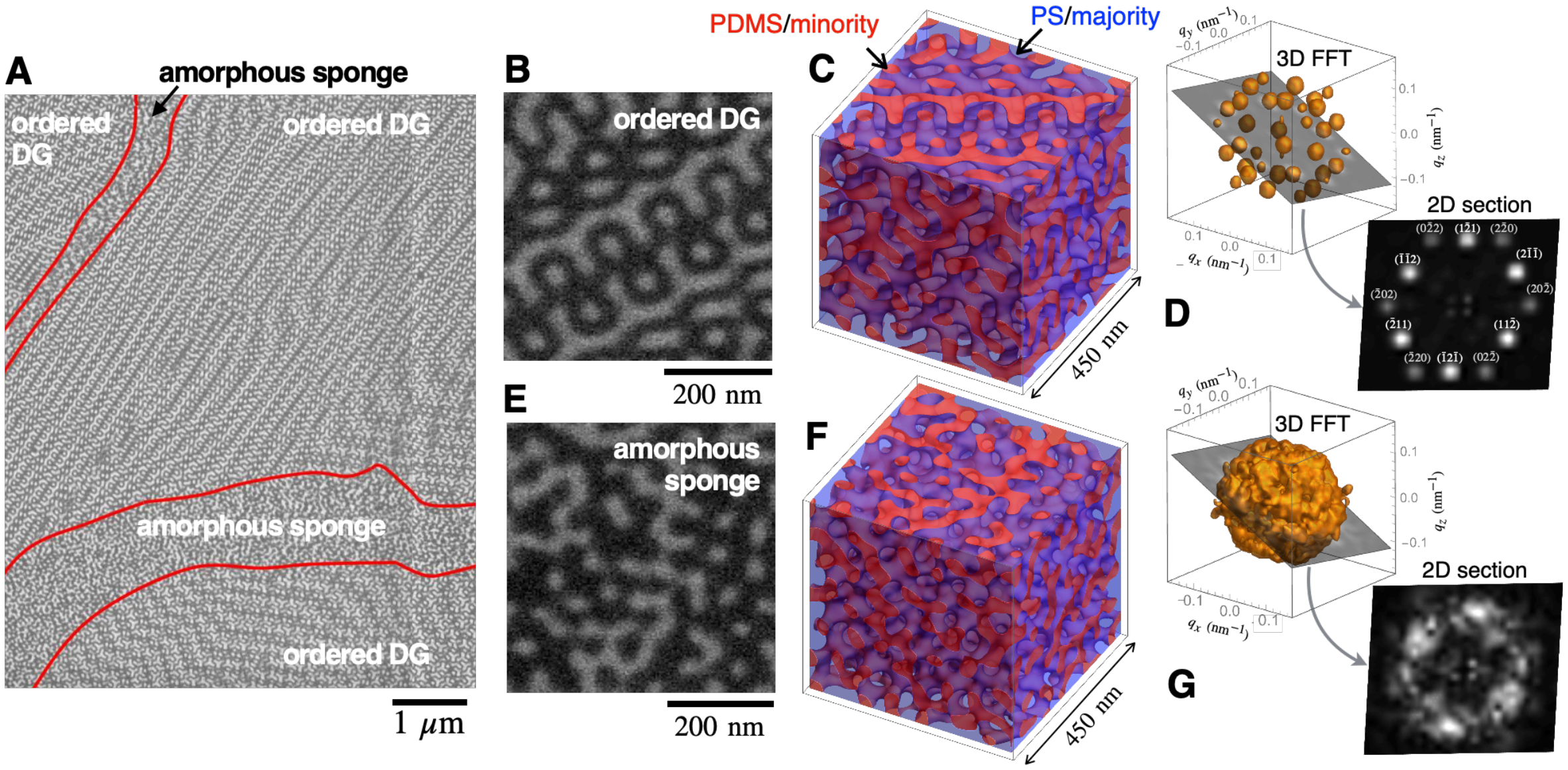

**Figure 1 – Coexistence between ordered double-network and amorphous sponge.** In (A), large field of view of 2D SEM section of the PS-PDMS sample, with highlighted regions of ordered DG (cPCN) and random sponge (aPCN) separated by boundaries highlighted in red. 2D SEM image sections of analyzed sample volumes of cPCN and aPCN are shown in (B) and (E), respectively. 3D renderings of a smaller subregion of the SVSEM reconstructions of ordered DG and amorphous sponge are shown in (C) and (F), respectively, with red and transparent blue indicating locations of PDMS and PS enriched regions, respectively. Fast Fourier Transform (FFT) analyses of the respective 3D volumes are shown in (D) and (G) with the top showing a 3D density plots of intensity (linear scale) in 3D reciprocal space, with large intensity at wavevectors near masked out for visibility of finite-$q$ density correlations. 2D slices are shown through the planes highlighted, with the section of (D) corresponding to a (111) plane of the DG, with indexed peaks as labeled.

## Compositional correlations & network structure

Renderings of 3D reconstructed volumes of aPCN and DG regions are shown in **Fig. 1C,F**, with the PDMS (minority) component and the PS (majority) component colored respectively as red and semi-transparent blue. The repeating domain pattern on the cubic faces of the sample volume for the DG region are consistent with 3D periodicity. The 3D Fourier transform of PDMS density in this region in **Fig. 1D** shows Bragg spots consistent with crystalline order of the domains. As describe previously, this order corresponds to a triclinically distorted variant of the ideal cubic ($Ia\overline{3}d$) structure of DG, the most prominent Fourier intensity is from a family of {211} peaks, whose averaged wavevector is close to the peak of the orientationally-averaged intensity at $q^*$ = 0.13 $nm^{-1}$, corresponding to cubic lattice parameters of roughly 120 nm. The domain patterns in the amorphous sponge lack obvious periodicity, consistent with the Fourier transform of the structure (**Fig. 1F**), showing intensity in a somewhat speckled halo peaked around wavevector $q^*$ = 0.11 $nm^{-1}$. While not isotropic due to the finite size of the FFT volume, the structure lacks strong signatures of anisotropic orientational and positional (i.e. crystalline) order apparent for DG. In **Fig. S1**, we compare the peak scattering wave-vectors from radially integrated profile of intensity versus wavevector. While for crystalline structures the peak wavevector $q^*$ has complex relationship between local neighbor "spacing" (due to reciprocal lattice selection rules), for aPCN structures, this peak scattering is often used to quantify a "domain size" or *D*-spacing, $D = 2\pi/q^*$, which roughly corresponds to the characteristic center-to-center distance between like-material regions. Notwithstanding the ambiguities implicit in the comparison of apparent peak-scattering positions between crystalline and amorphous structures, we note that this measure suggests somewhat larger *D* for amorphous (55 nm) than for ordered DG (48 nm)[1]. Below, we employ direct real-space structural analysis to resolve the inconsistency of this size comparison between ordered and disordered structures, and show how different measures of relative size of the two morphologies derive from distinct features of their molecular packing.

We next compare the geometry of networks that thread through the tubular PDMS domains by thinning these domains down to 1D graphs of nodes connected by straight edges, called struts, as shown in **Figure 2**. As described previously (20, 46), this approach places nodes and struts of these skeletal graphs at positions of maximal PDMS density, presumably corresponding to the "central axes" of the tubular domains and nodes at the centers of inter-tubular junctions. See **Supplementary Video 1** for animated views of skeletons from both regions. For the ordered DG, the network is composed of three-coordinated nodes, which are situated at positions associated with two gyroid networks (shown in **Fig. 2B**)— known alternately as (10,3)-a or *srs* nets— somewhat distorted from perfect cubic symmetry (20, 21). Accordingly, as shown in **Fig. 2C** the internal node angles $\theta$ at the nodes are distributed around a mean value $\langle\theta\rangle = 119.7^\circ \pm 7.4^\circ$ consistent with the nearly planar and three-fold geometry of gyroid nodes. Additionally, we characterize the local dihedral angles $\psi \in [0°, 180°]$ between consecutive triplets of skeleton edges (46), and categorize the chirality of edges as right- or left-handed for $\psi < 90^\circ$ and $\psi >$ $90^{\circ}$respectively **Figure 2A** shows PDMS network edges colored according to the mean $\psi$ value, clearly splitting the two gyroid subgraphs as uniformly right- or left-handed (**Fig. 2B**), close to the values of the enantiomeric twist of the cubic DG, $\psi(\text{right}) = 70.5^{\circ}$ and $\psi(\text{left}) = 109.5^{\circ}$ [2].

---

[1] Strictly speaking, a strong peak (narrow and well separated) in orientationally averaged Fourier intensity at $q^*$ implies a peak in orientationally averaged pair-correlations (between like-component segments) at a characteristic distance proportional to $D = 2\pi/q^*$, (44, 45).

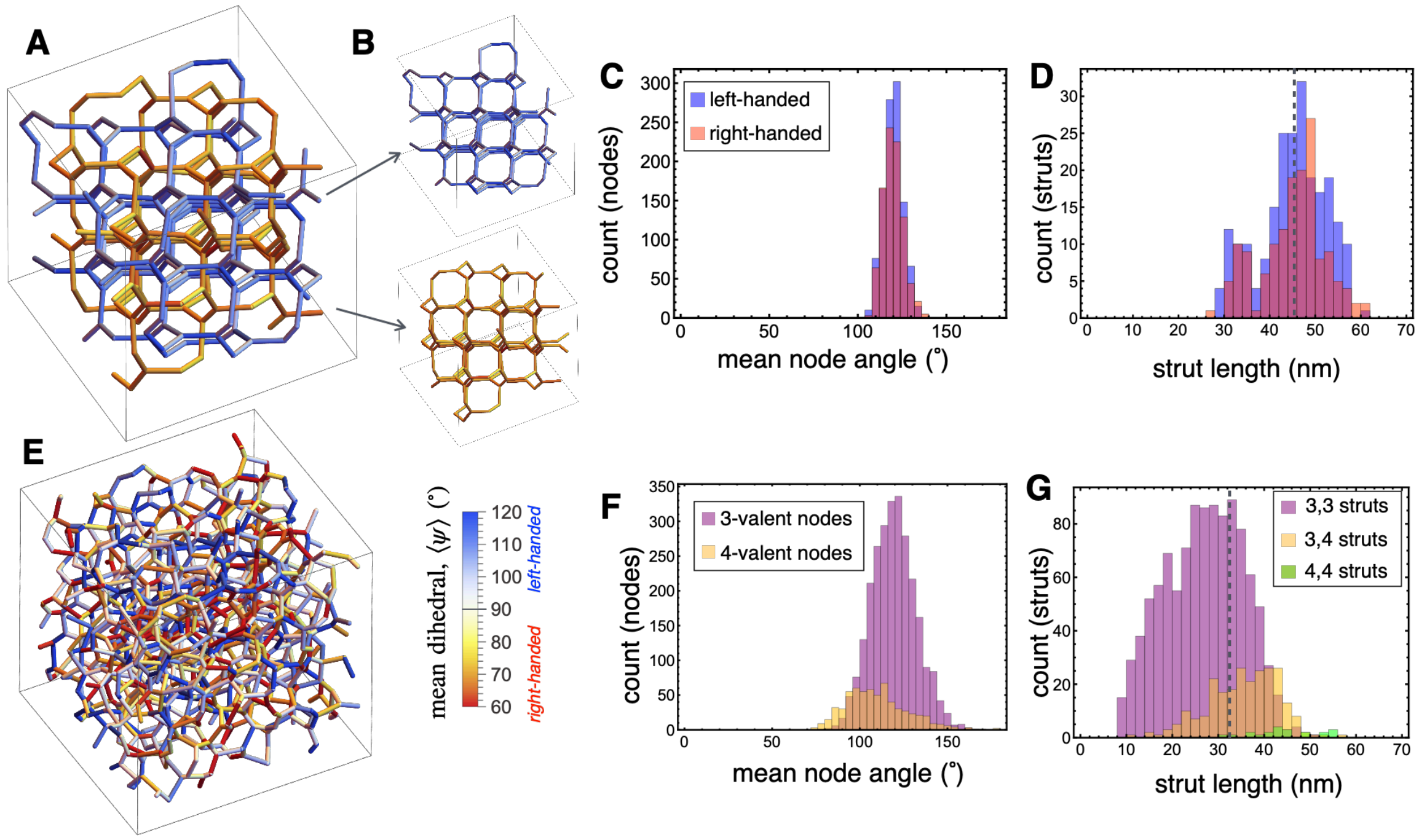

**Figure 2 – Comparing geometry between ordered and disordered networks.** The skeletal graphs of reconstructed ordered DG and sponge are shown respectively in (A) and (E), with struts shaded according to the local mean value of dihedral twist at each edge. In (B), the left- (blue) and right-handed (red) single gyroid skeletons are shown as separated networks. Histograms of internal node angles are plotted for ordered DG and sponge in (C) and (F), respectively, and corresponding strut length histograms are shown in (D) and (G), with colors in (D) same as in (C). For ordered DG, histograms from right- and left-handed skeletal graphs are overlaid, while for sponge, distributions for different node valence and struts connecting different node valence are overlaid. The dashed vertical line indicates the mean strut length in (D) and (G) for each structure.

By contrast, the PDMS network of the sponge consists of a single connected graph that is considerably more heterogeneous by the same measure (**Fig. 2E**). While the vast majority (91%) of nodes are trivalent ($z$ = 3), the network is composed of a significant subpopulation (9%) of tetravalent ($z$ = 4) nodes. The internal angle $\theta$ distributions of node types are shown in **Fig. 2F**. While $z$ = 3 nodes exhibit a mean angle $\langle\theta\rangle_{z=3} = 118.6^{\circ} \pm 12.7^{\circ}$ corresponding, on average, to planar-trihedral geometry, these angles are spread over a considerably broader range in comparison to ordered DG nodes, characterized by 70% larger standard deviation. Likewise, $z$ = 4 nodes exhibit internal angles that are distributed around a mean value $\langle\theta\rangle_{z=4} = 109.2^{\circ} \pm 17.5^{\circ}$ that is close to the tetrahedral symmetry, although angular fluctuations are even broader than in the trivalent nodes. In contrast to the coherent inter-node chirality with the two gyroid networks of the ordered region, the PDMS network exhibits strong local fluctuations in the chirality of dihedral twist as measured by maps of $\Delta\psi = \psi - 90^{\circ}$, shown in **Fig. 2E**, which exhibits alternation between right-handed and left-handed twist *within the same single network.*

---

[2] We follow a prior convention to assign handedness based on the right- or left-handed sense of 70.5° rotation of trihedral plane between adjacent nodes (20, 46).

While edge chiralities might be anticipated to be purely random in the amorphous network, an analysis of the chirality correlations $\langle \Delta\psi_i \, \Delta\psi_j \rangle$ for different strut pairs *i* and *j* shows that these are spatially correlated, plotted in **Fig. S2A** as function of graph distance between struts pairs *i* and *j*. In contrast to the long-range correlations within the ordered DG networks, the chirality in the amorphous sponge exhibits only short-range correlations that decays away at distances beyond a few (~3-4) edge lengths. This means that while chirality is disordered and, on average, enantiomeric on large-scales in the amorphous PDMS network, there exist spatial zones of like-chirality of a size scale comparable to multiple internodal distances, shown illustrated in 3D as extended, string-like regions of chirality in **Fig. S2B-C**.

The distributions of strut lengths between neighbor nodes of PDMS networks in both the ordered DG and amorphous sponge regions are compared in **Fig. 2D and G**, respectively. As described previously, the ordered PS-PDMS DG exhibits a non-affine, triclinic distortion of the cubic double network, leading to families of relatively shorter or longer struts grouped according to their orientations (20, 21). Indeed, this coherent non-affine pattern of relatively long vs. shorter is the dominant origin of edge length variability in the ordered DG networks, which have a mean length of 44 nm. By comparison, edges of amorphous networks are more intrinsically variable in length, independent of the node valence they connect. The mean length (29 nm) of amorphous network edges is substantially shorter than in the ordered DG networks. Notably, this trend is opposite to relative size scale measured peak Fourier intensity, demonstrating that differences in relative apparent *D*-spacing do *not* correlate with differences between intra-network spacing of neighbor nodal units.

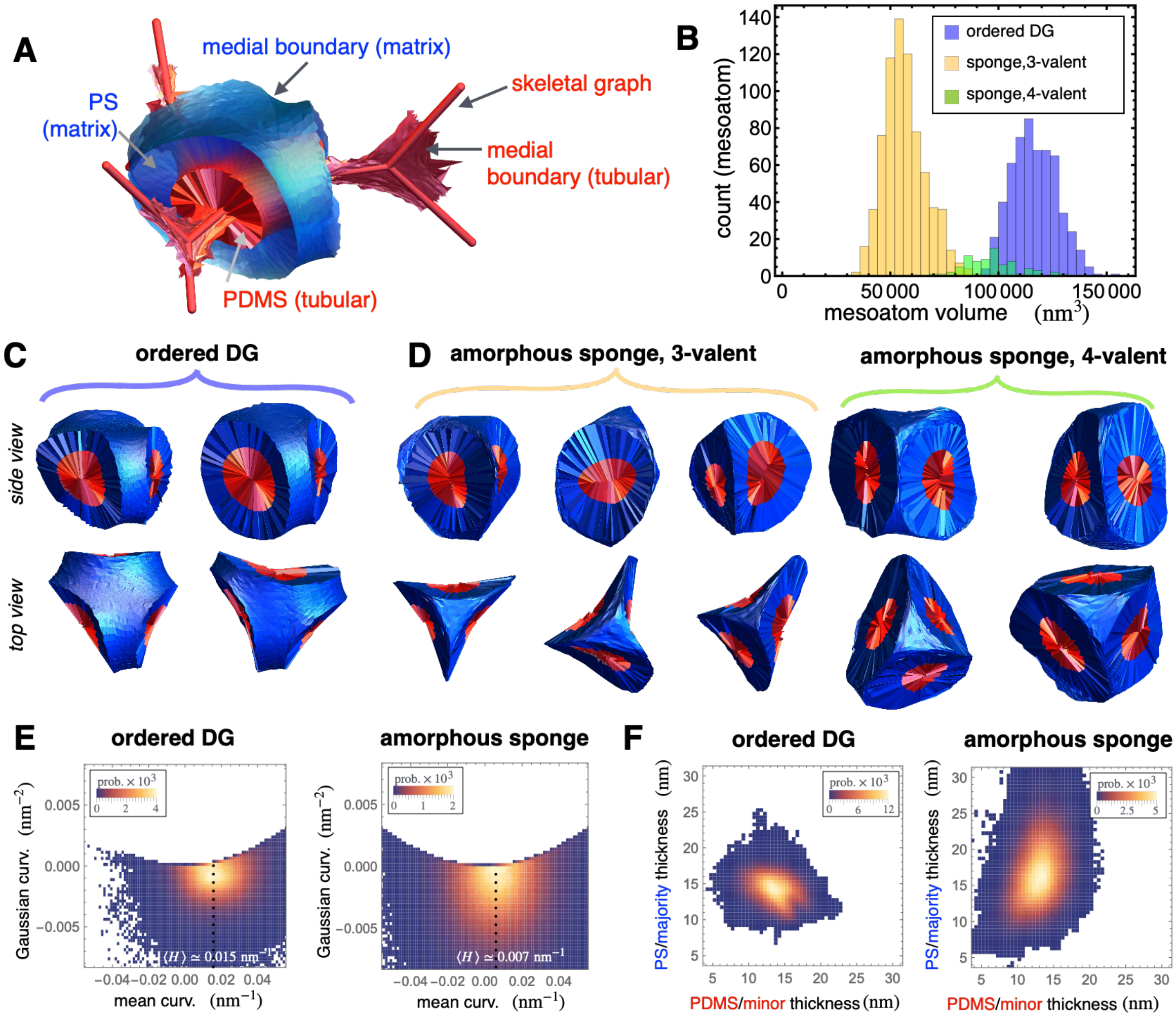

**Figure 3 – Comparing mesoatomic shapes and macromolecular packing geometry.** In (A), the medial geometry surrounding a node of DG, defining the mesoatomic unit of the cPCN. For clarity, the PS (matrix) region is shown as semi-transparent, but spans from the IMDS to the medial boundary in the center of the PS subdomain, while for visibility the web-like medial boundary of the PDMS phase is shown extending out in the neighbor nodes to the highlighted mesoatom. Histograms of the mesoatom volumes of ordered DG and sponge are shown (B), with 3D renderings of examples mesoatoms are shown for both structures in (C) and (D). 2D histograms of IDMS curvature as functions of mean and Gaussian curvature are compared in (E), with the dashed vertical line highlighting average mean curvature value for each. 2D histograms of local medial thickness as functions of PDMS and PS subdomain thickness are compared in (F) for both ordered DG and sponge.

## Mesoatomic motifs and local domain geometry

We now turn to consider the local domain geometry, first focusing on the shapes and features of the nodal mesoatoms. As introduced elsewhere, mesoatoms correspond to molecular groupings that constitute geometric "building blocks" of the amphiphilic assembly. For PCN assemblies, it is especially useful to consider the molecular groupings associated with the nodal interconnections to define mesoatoms. SVSEM does not directly image individual macromolecules but instead reveals the boundary between PS and PDMS rich regions. Nevertheless, we may use the medial geometry (5, 47, 48) of the IMDS between PS-PDMS as a proxy for where the ends of the brush-like groups of polymer blocks contact within each minor and major subdomain, as shown for an example in **Fig. 3A**. This medial analysis voxelates the sample into narrow wedge-like volumes that capture the set of positions (volumes) closest to each given point on the IMDS, providing a proxy for the optimal (i.e. maximal entropy) polymer chain configurations that pass through each IMDS point (see Methods). As described previously (19, 49), for DG, these medial sets correspond to a smooth, Gyroid TMPS-like surface through the middle of the matrix subdomain and to webs of twisted ribbon-like surfaces stretching over the skeletal graph of the minority block tubular network subdomains. By selecting the groupings of medial wedges that are closest to a given network node, we arrive at the derived mesoatom for that node, shown for an example node from the ordered DG region in **Fig. 3A**. See **Supplementary Video 2** for animated views of some example mesoatoms from DG and sponge.

Comparing the shapes of the several different mesoatoms from the ordered DG in **Fig. 3C** , some shape variability is clear, yet each shares features of the non-convex shape of the "ideal" mesoatomic unit of the cubic DG (15). There are three, roughly planar, "strut face" boundaries containing both components contacting the nearest mesoatomic neighbors located along <110> directions in the same network for DG. The remainder of the mesoatom boundary consists of a smooth saddle-like surface comprised of the majority block. Additionally, each DG mesoatom makes contact with 14 neighbor mesoatoms from the opposite chirality network across the smooth, negatively-curved outer boundary, what we call a "saddle-skin" motif. There are 2 next nearest neighbors stacked along the 3-fold monkey saddles "above" and "below" the plane of the trihedral node and the other 12 next-next nearest neighbors contact along 3 "elbow" regions that span from the top to bottom monkey-saddle regions in the region between adjacent strut pairs.

As shown in examples in **Fig. 3D**, the mesoatoms of the sponge, fall into two categories: trivalent (91%) and tetravalent (9%). The trivalent mesoatoms are different than the DG mesoatoms, since while they have three strut faces, each containing the central PDMS region and the surrounding PS, the only other well-defined (i.e. non-singular) mesoatom boundaries are the two roughly planar PS facing boundaries "above" and "below" the nominal trivalent plane. In contrast to the smooth, saddle-skin boundaries of the DG mesoatoms, the corresponding regions that span between strut faces of the amorphous mesoatoms contain rather sharp and cusped surfaces, appearing similar to an axeblade in shape on the outer mesoatom boundary. The tetravalent mesoatoms of the amorphous network are similarly reminiscent of the tetrahedral mesoatoms of the double-diamond morphology, again with the exception that the smooth elbow regions between adjacent mesoatoms are instead tapered into sharp axeblade-like geometries. We return to the geometric origin of these singular boundary cusps in the next section.

Comparing the volumes of the mesoatoms in **Fig 3B**, we find that, on average, mesoatoms of the ordered DG are roughly twice the mean volume of sponge mesoatoms. Based on the BCP composition and density of the PS and PDMS diblocks, we estimate that these correspond to volumes composed of roughly 860 and 450 chains respectively. The relatively larger volumes of mesoatoms in ordered DG in comparison to the amorphous sponge is consistent with the relatively longer lengths of the struts in their PDMS networks. Moreover, within the population of sponge mesoatoms, the volumes are heterogeneously distributed between different nodal valance, with tetravalent mesoatoms assuming a roughly 70% larger mean volume than the trivalent mesoatoms. This relatively smaller volume is consistent with the smaller mean strut length (**Fig. 2**) in the sponge relative to ordered DG, but again would seem to run counter to its apparently larger size as measured by *D*-spacing comparison above.

We next compare two geometric measures of chain packing within these mesoatomic motifs: IMDS curvature and block subdomain thickness distributions. **Fig. 3E** compares the 2D histograms of the mean and Gaussian curvatures of the IMDS. Notably, the ordered DG and disordered sponge display comparable variability of IDMS curvature values, notwithstanding obvious lack of long-range order of the latter structure. In comparison to the ordered DG morphology, the IMDS of the disordered sponge skews to more strongly negative values of Gaussian curvature and to lower values (less positive) of mean curvature, roughly half as large on average. In **Fig. 3F**, we compare the 2D histograms of major (PS) and minor (PDMS) of block subdomain thickness extracted from the medial analysis for ordered DG and amorphous sponge. The distributions of PDMS block thickness are surprisingly similar for both cases, both in terms of the average value of approximately 13 nm and the variability, suggesting the interior packing of the tubular domains of the networks is rather similar in ordered DG and amorphous sponge. In contrast, the majority PS blocks are skewed to larger thickness for the amorphous mesoatoms relative to ordered DG, corresponding to mean block sizes of 17 nm and 14 nm, respectively. This statistically significant offset in PS block thickness suggests that the major disruption to the chain packing due to the random connections of the mostly trivalent and similarly shaped junctions takes place on the "outside" of the tubular PDMS domain for sponge mesoatoms.  Hence, while mesoatomic units of sponge can be *smaller* than their DG counterparts by measures of their strut length and volume, they are in fact 11% *larger* by measure of their domain thickness, due to the relative extension of the major PS block in the amorphous structure. We next show that this distinction in packing on the chain scale derives from the structural features of the networks at the scale of loops.

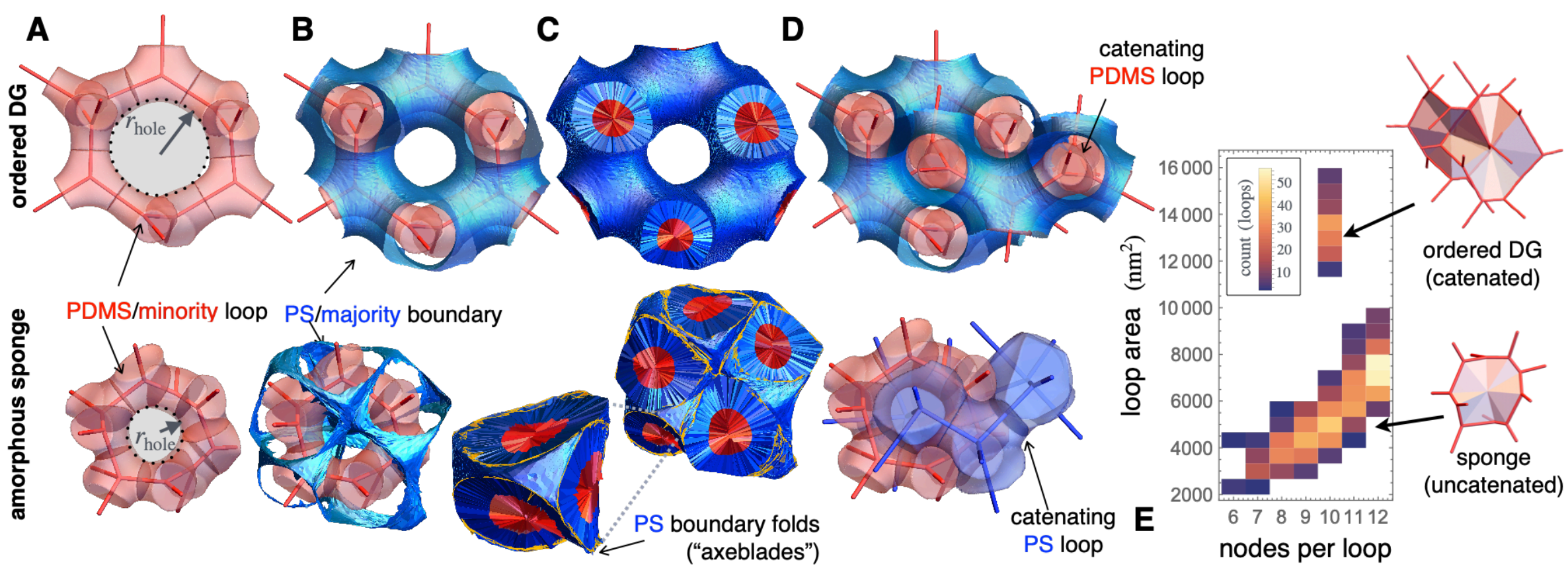

**Figure 4 – Loops and links in ordered and disordered network.** (A) –(D) Comparison of example network loops from ordered DG (top) and amorphous sponge (bottom), consisting of 10 and 8 nodes respectively. (A) shows PDMS subdomains and skeletal graphs, with the grey circular region highlighting the different apparent radii of the interior holes at the center of the PDMS loops. (B) Shows the outer (medial) PS boundary of the mesoatoms belonging to the loop as transparent blue, and (C) shows the volume filling rendering of the PDMS (red) and (PS) subdomains within the loops. Notably, the PS region fills the interior of the sponge loop, leading to sharp folds ("axeblades") in PS boundaries where multiple mesoatoms converge at the center of the filled loop (highlighted as yellow facets). (D) shows the DG loop catenated by one of the loops of the other PDMS network, while the sponge loops are not catenated by another PDMS loop, but are instead catenated by loops of the tubular network composed by PS. (E) shows a 2D histogram of loops of ordered DG and amorphous sponge in terms of their node number and area spanned (triangulated skew polygon shown in pink). All loops of DG are size 10, and are catenated by PDMS loops of the other network, and therefore span larger areas than loops of sponge, which vary in node number, but are all *uncatenated* by other PDMS loops.

## Connecting loops and links to molecular packing

To understand the distinctions in chain packing geometry, we analyze and compare the structure of PDMS network loops in the ordered and disordered PCN assemblies. See **Supplementary Video 3** for comparative animations example loops from DG and sponge structures. In the ordered DG structure, as shown in **Fig. 4A** (top), loops are uniformly composed of 10 trivalent junctions of tubular PDMS domains. In **Fig. 4B** (top) we render the outer boundary PS (matrix) domain as well as the filled domain volumes of the mesoatoms in the 10-loop, which shows clearly an interior "hole" through the center of the loop, indicating that chains associated with the loop nodes do not fill this core zone. Instead, the holes through each 10-loop of the DG are threaded through by loops from the other (enantiomeric) PDMS resulting in the double-network morphology. As shown in **Fig. 4C,D** (top), the hole interior to the PDMS loop is filled by its associated PS block in combination with the PS and PDMS domains in struts for the other network that pass through the loop, maintaining smooth contact between mesoatoms along the PS saddle-skin boundary. We directly confirm that all loops in ordered DG (not including loops that are cut by the sample boundary) are catenated by loops from the opposite single gyroid network.

In contrast to the ordered DG, loops of the sponge network are composed of variable numbers of edges and, in general, have mixtures of tri- and tetra-valent mesoatoms. We illustrate an example 9-loop composed of 8 trivalent nodes and 1 tetravalent node in **Fig. 4A** (bottom). The apparent radius of the hole that threads through the tubular PDMS domains is roughly half that of the hole through the 10-loop of the ordered network. Since the thickness of the tubular PDMS domains is essentially unchanged between the ordered and amorphous PCNs, the smaller hole radius explains the shift to lower mean curvature and more negative Gaussian curvature for the latter morphology (much like the effect on the inner part of a torus with fixed minor radius and a shrinking major radius), as observed in **Fig. 3**. The smaller average diameter of amorphous network loops is also consistent with the shorter strut lengths measured in **Fig. 2**, notwithstanding the fact that sponge loops have variable number of nodes with as few as 6.

More significantly, the smaller diameter of sponge loops further implies that there is not sufficient room to fit another tubular domain threading through the center of the loop, such as occurs for the ordered DG loops. This is clear when rendering the outer boundary of the PS (majority) as well as the filled mesoatom volumes of the loop in Fig. **4B-C** (bottom), which show that the PS domains belonging to the loop mesoatoms fully plug the hole. This PS-filled center has the direct consequence that the outer mesoatom boundaries meet at sharp folds (highlighted in yellow), clearly unlike the smooth PS domain contact along the gyroid-like surface in the ordered DG. Geometrically, these folded boundaries arise due to the constraints of filling all points within the hole with chains from the loop mesoatoms, leading to a wedge-like geometry akin to the interior corners of a Voronoi polyhedron. Indeed, the fact that PS blocks must stretch to fill the angular corners at the center of the uncatenated loops is the direct cause of the relatively larger and broader extension of PS blocks in amorphous sponge than in ordered DG. In **Fig. S3** we show additional examples of PDMS loops of varying size from 6 to 11, highlight similar features of PS boundary creasing due to the packing at the loop interior.

Considering all loops (up to length 12) in both sample regions, we confirm that every loop of DG is catenated by loops of the other network, while no loops of the amorphous networks we analyzed are threaded or linked by another tubular PDMS domain. In **Fig . 4E**, we show that this contrast between linked and unlinked directly correlates the area spanned by the loops in the different regions, with spanned areas of ordered DG loops roughly twice that of the sponge loops possessing the same and even large numbers of edges (i.e. 10 or 11 nodes).

From this perspective, we observe that the sponge is in fact a type of *single-network* morphology, and in terms of nodal composition, most similar to a single gyroid structure, albeit randomly rearranged and intermixed with a small fraction of diamond-like nodes. This view of the amorphous sponge as a random, mostly trivalent structure implies a crucial difference in topology of the PS domain from that of the ordered DG, where PS blocks form a thickened G TPMS-like matrix separating the two disjoint tubular PDMS networks. In the case of the sponge, the single network character suggests that the PS domain itself constitutes a second tubular network domain, which is by definition disjoint from the PDMS network but with both the PDMS and PS networks mutually intercatenating. In **Figure 4D** (bottom), we show an example of a so-constructed PS loop that catenates the example 8-loop of PDMS mesoatoms from the amorphous regions. While we leave the detailed analysis of this PS network, and its relationship to the PDMS network, for future work, we note only briefly (see **Fig. S4**) that it is composed of a different mixture of node types – 75% trivalent, 19% tetravalent, 6% pentavalent – which could reflect differences due to the fact that PDMS and PS occupy different volume fractions.

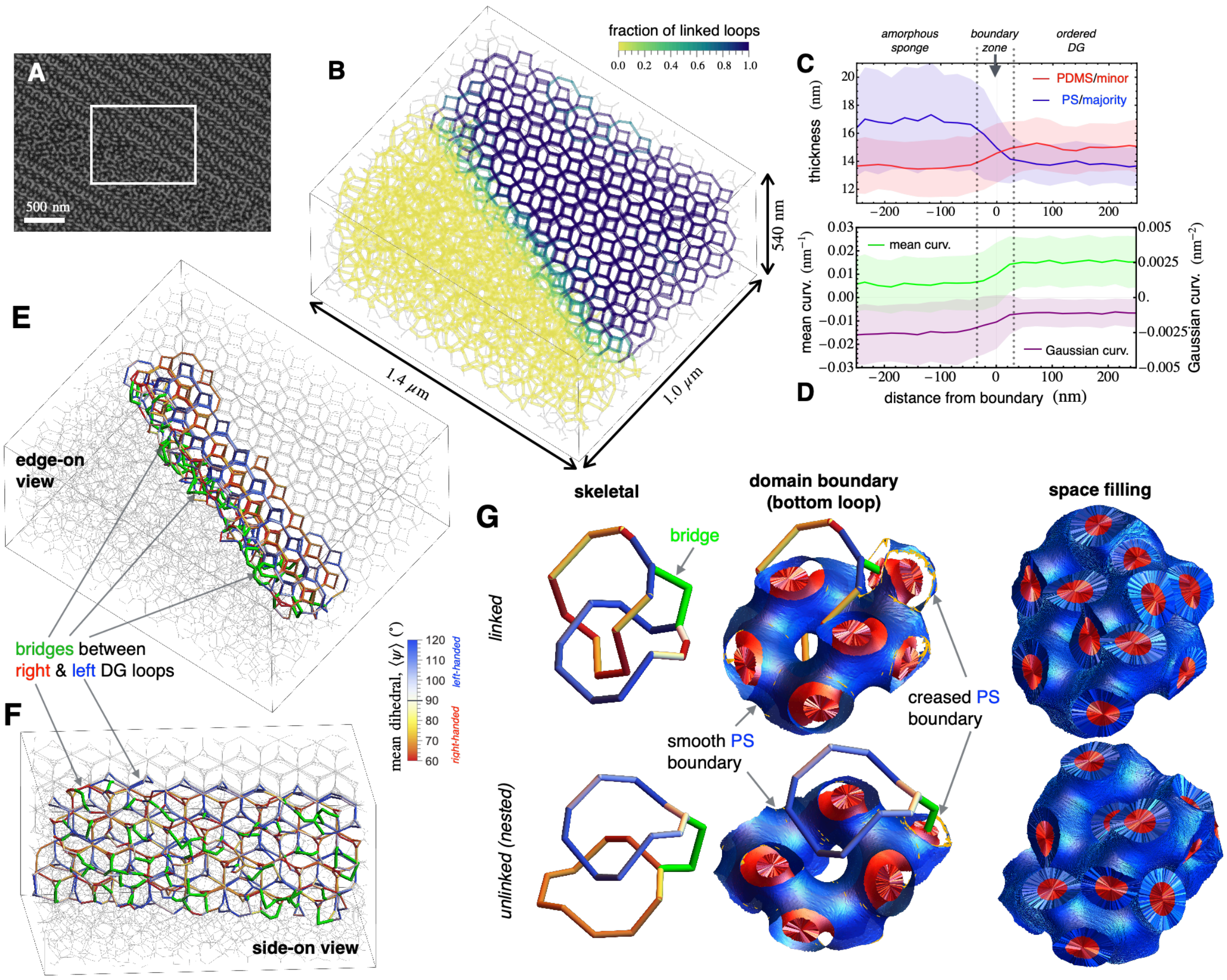

**Figure 5 – Boundary between ordered and disordered network morphology.** (A) 2D SEM image showing the analyzed boundary region between ordered DG and amorphous sponge. (B) shows a skeleton of the PDMS network extracted from SVSEM imaging in boundary region, with struts colored according to local linking fraction, i.e. fraction of loops belonging to a given strut that are linked by other loops (up to size 12), partitioning the skeleton into fully linked (dark blue) and fully unlinked (yellow), in the respective DG and sponge regions, separated by a narrow band of partially linked edges. (C) and (D) show analyses of local block thickness and IMDS curvature, indicating that the transition molecular packing geometry takes place over a narrow zone of size comparable to the dimensions of a single mesoatom (~50 nm). (E) and (F) show illustrations of the “short-circuits” (green struts) between opposite chirality networks of ordered DG at the boundary with amorphous sponge. (G) shows two examples of short-circuiting loop geometries, with PDMS domains and PS boundaries rendered for the bottom loop in the middle column and space-filling mesoatoms rendered in the right column. Notably, sharp creased PS boundaries (“axeblades”) appear at the edges of short circuiting mesoatoms.

## Where order meets disorder

Having established the absence of PDMS loop linking as a primary distinguishing feature of the amorphous sponge network, we turn to the specimen regions at the boundary between the ordered and disordered morphologies. The highlighted region of the SEM image in **Fig. 5A** has been reconstructed: the skeletal network of the PDMS subdomain is shown in **Figure 5B**, with edges colored according to the local *linking fraction*, which counts the fraction of loops that pass through a given edge that are linked by other loops (counting loops up to size 12). Notably this splits the region neatly into predominantly fully linked (dark blue) or fully unlinked (yellow) regions, with the former showing obvious DG order from a <100> viewing direction, while the latter is clearly spatially disordered. A relatively narrow region of partially linked edges separates the fully-linked, ordered DG and fully-unlinked amorphous sponge.

Approximating the boundary of the fully-linked region (DG) by its convex hull, we sample the measures of the local domain geometry (curvatures and thickness) as function of distance from the order-disorder boundary and plot these in **Fig. 5C** and **D**. These show that average local thickness and IMDS curvatures transition between the characteristic values analyzed in **Fig. 5** for amorphous sponge relative to ordered DG across a relatively narrow boundary zone. The width of this zone, roughly 50-60 nm, is comparable to the dimensions of the mesatomic nodal units of the networks themselves. This implies a fairly abrupt boundary between these two morphologies notwithstanding the stark contrasts between their long-range order, symmetry, and network topology. We show a similar spatial profile in the transition of edge lengths and mesoatom volumes in **Fig. S5**. See **Supplementary Video 4** for animated view of the set of mesoatoms at this boundary, viewed from both the ordered DG and disordered sponge regions.

The sharp boundary between these structures poses a puzzle about how the morphology transforms from a single network of linked loops on one side of the boundary to an interlinked enantiomorphic double-network just on the other side of that boundary. That is, at this boundary the disjoint and opposite chirality networks of DG must effectively emerge from the single and uncatenated network of the amorphous sponge. Notably, this rather narrow boundary zone is in direct contrast to the relatively broad gradient transition zone observed recently for order-to-order transformations between DG and double-diamond in PS-PDMS melts (50). To illuminate the local structure of this 1-to-2 network transformation, we extract the "short-circuiting" *bridges* from between the right- and left-handed networks of the DG, and show that these (green edges) localize at the frontier between order and disordered network regions in **Figs. 5E-F**. In brief, bridges are identified by extracting pairs of linked loops with opposite chirality that are connected by extra edges (i.e. that are absent in the DG network), considering the shortest possible graph distance. We consider bridges of up to length 5 extra edges, although the overwhelming majority are found to be length 1 or 2 extra edges.

Analyzing the bridging geometry further, we see that these short circuits fall into two basic motifs, as shown in **Fig. 5G**: bridges between interlinked or non-linked (but nested) neighbor loops. Visualizing the mesoatomic shapes of those loops, it is obvious that the linked and chiral loops have smooth and continuous PS boundaries characteristic of DG mesoatoms. In contrast, the bridges themselves are composed of mesoatoms with sharp PS boundaries, characteristic of the amorphous sponge and the requisite packing frustration at the center of its uncatenated loops. Below we discuss the potential role of these bridges in the putative mechanism of ordered DG growth at the front of this boundary with the amorphous sponge.

## Discussion

In summary, direct 3D reconstruction of the amorphous sponge morphology of PS-PDMS reveals a disordered, single-gyroid like morphology. That is, the minor component (PMDS) constitutes only a single network whose loops are not linked by other parts of the PMDS network, but instead linked by an effectively tubular network of the major component (PS). While it lacks long-range order and regular loop distribution of a proper single-gyroid, we consider it *gyroidal* in the sense that its nodes are mostly 3-valent and planar, albeit with a small fraction of diamond-like (i.e. 4-valent) nodes.

The precise mechanisms under which the disordered sponge forms in this PS-PDMS system remains to be understood. As equilibrium morphologies are widely predicted to have long-range order at the well-segregated conditions studied here, we interpret the appearance of the amorphous single-gyroid sponge as a kinetically-trapped, non-equilibrium precursor to the more thermodynamically preferred ordered DG, which becomes kinetically frozen as solvent evaporates from the sample and the PS regions vitrify. The disordered sponge structure likely forms rapidly, either due to assembly of aggregated chains in solvent, or else via a spinodal-like growth of composition fluctuations, and becomes kinetically frozen when the PS regions vitrify. Previously (15), we proposed one scenario in which an ordered DG structures forms by the nucleation and growth from pre-assembled, micelle-like primordial mesoatoms, first forming a small cluster of potentially poorly-ordered mesoatoms that later evolves to a double network with addition of additional mesoatoms into a faceted cubosome-like structure. Notwithstanding the unknown kinetic mechanism underlying the single-network sponge observed here, we speculate that if the PS component were provided sufficient mobility and time, this disordered single-network would transform to the ordered double-network. Likely, DG grains nucleate and grow from the primordial sponge, such that at longer times the size of the amorphous inter-grain zones shrink, and eventually disappear from a fully ordered sample.

Although kinetics and structural mechanisms of the disorder to order transformations clearly necessitate further study (e.g. by time-resolved observations of relative amorphous/ordered fractions and structural evolution from primordial sponge mesoatoms to mature DG mesoatoms), the idea that a disordered single-network morphology may be a precursor to ordered BCP phase is anticipated by prior work by Hashimoto and coworkers, in which ordered lamellar grains are observed to nucleate and grow from a disordered, apparently bicontinuous morphology (51). Additionally, Yadav and coworkers have used self-consistent field calculations to show that the disordered state of weakly-segregated AB diblock copolymers is well-modeled by a single-gyroid morphology, as opposed to the double-gyroid which becomes equilibrium above the order-disorder transition (52). We note, however, that in our study, the PS-PDMS is strongly-segregated and therefore deep into the regime of thermodynamically stable ordered states, which suggests further that the disordered network persists as an at least somewhat metastable state in this region. In this sense, the disordered single-gyroid may be the “network liquid” analog to the “liquid-like” micelle state of disordered spherical domains which has been proposed as structural model of the observed disordered precursors to ordered complex crystals, like the Frank-Kasper phases (53–56). Notably, a similar picture is proposed for the blue phases (57)—tubular networks of double-twisted texture in thermotropic chiral liquid crystals— which exhibit two canonical crystalline equilibrium phases (BP I and II), as well as third disordered network (BP III) that is argued to be favored by entropic effects (58, 59).

The structural correlations of the amorphous gyroid sponge observed here raise a number of key questions. First, assuming this disordered network morphology to be a metastable precursor to a thermodynamically favorable ordered double-network, what are the local mechanisms of transformation between these states? Naïvely, since the mesoatomic units of the amorphous network are largely 3-coordinated, like the DG, it might be expected that simple rotations and reconnections of these units occurring at the boundary might be sufficient to convert the single, uncatenated network into a double, intercatenated network. However, the fact that 3-valent mesoatoms in the ordered DG are approximately twice as large in volume compared to their 3-valent counterparts in the sponge implies that the spatial density of nodes is lower in the former state. Accordingly, one might expect that roughly 2 sponge mesoatoms could simply fuse to form a single DG node. However, this would require that the average valence of nodes to increase, while the mean valence in DG is slightly lower than in sponge. This implies that in the transformation from sponge to DG, "excess" struts of the amorphous network must be "pruned" as the frontier of ordered DG expands into the disordered region, since cutting struts between 3-valent nodes simply reduces the total node count by 2 (and increases the mean length between the remaining struts). This, indeed, may point to *scission* of the short-circuiting struts between the otherwise disjoint single-gyroid network (as extracted in Fig. 5) as the key dynamical mechanism of disorder to order transformation.

On a more mathematical note, the Gyroid surface that divides the DG into two enantiomeric networks is oriented: its normal vector is defined globally. The disordered region connects the two networks together, ruining the global orientation of the surface. Möbius defects have been proposed in lyotropic bicontinuous phases previously (60), and the "hole punch" structures in the medial PS surface appearing in short-circuiting bridges between opposite chirality DG loops at the boundary with the sponge (e.g. **Fig. 5G**) may be the first observed example of this.

Beyond its transformation path to a more ordered state, a more basic question remains about the disordered network morphology: why is the amorphous sponge a single-network type morphology? Presumably, the extra packing frustration in the PS domains at the center of the uncatenated PDMS loops leads to a higher thermodynamic cost (19), and metastabilty, relative to the ordered network. If so, then it is all the more puzzling why the PDMS loops are uniformly uncatenated in the amorphous sponge. Is there a mechanism to prevent at least some of these loops from random threading by another part of the PDMS network? Is the lack of catenation a feature of particularly high configurational entropy in random networks? We note that several simulation models of associating systems that form ordered networks like gyroid (61) and diamond (62) show at least metastable, if not equilibrium, disordered network liquids (62, 63). Indeed, some such models have been shown to exhibit low-density and high-density states distinguished by the nature of unlinked versus linked random loops (64, 65). Or else, the lack of catenated loops may be a feature of the nature of "spongey" density fluctuations from which the initial network is formed. In that case, it may be expected that the structural correlations of loops and their linking properties in the disordered BCP sponge fall into the class of spinodal-like structures described by random Gaussian wave models (35, 66–68), although to our knowledge the statistical structure of such networks (loops, nodes and link distributions) is not known. Looking beyond the widely-studied models of spinodal structure, the connection of amorphous sponge morphologies to aperiodic minimal surface geometry (69) as well as a broader class of random-continuous network models (70, 71) remains to be explored.

Understanding what controls these structural correlations of polycontinuous random networks in BCP and other systems may open the door to engineering new and useful composite materials with tunable topologies and tortuosity, but without long-range order. The functional properties that may derive from correlated disorder of network materials (72–75), such as isotropic photonic band gaps, make them attractive targets for applications. Much like well-defined randomness of disordered particles and grains (76, 77), the random network materials likely offer similar opportunities to understand and harness their hidden topological order.

**Acknowledgements**

We gratefully acknowledge A. Avgeropoulos and G. Manesi for synthesis of the PS-PDMS diblock copolymer. We are also grateful to R. Hayward for valuable discussions of this work. This research was supported by the U.S. Department of Energy (DOE), Office of Basic Sciences, Division of Materials Sciences and Engineering, under award DE SC-0022229. Additional support was provided by the National Science Foundation under award DMR 2105296 (E.L.T. and X.F.) for support of SVSEM imaging.

## Methods

**Material and sample preparation.** Sequential anionic polymerization of styrene followed by hexamethylcyclotrisiloxane was employed to make the polystyrene-b-polydimethylsiloxane (PS-PDMS) diblock copolymer. The PS block was 50 kDa and the PDMS block 33 kDa with a PDMS volume fraction of 0.41 and a polydispersity of 1.05. The sample was dissolved at 10 wt % in toluene and the solvent slowly evaporated over the course of 7 days at room temperature, followed by vacuum annealing at 60 C for 3 days to assure all residual solvent was removed.

**Slice-and-view SEM data acquisition.** SVSEM data was acquired using a ThermoFisher Helios NanoLab G4 CX SEM-FIB. This dual electron beam – ion beam instrument uses a 30 keV gallium ion beam to mill slices from the sample surface. The sample surface was then sequentially imaged after ion beam slicing using a 1 keV electron beam and a through lens secondary electron detector. Due to the stronger scattering from the Si and O atoms in the PDMS block, the secondary electron signal is much higher (brighter) than for PS. The ion beam was also used to mill fiducials for registration of the successive images. The x-y pixel resolution was approximately 3 x 3 nm with a slice thickness of about 3 nm, resulting in a voxel resolution of $(3\ \mathrm{nm})^3$. Before performing SVSEM, the outer sample surface was coated with 50 nm of Pt for protection from ion milling beam damage.

**3D FFT analysis.** Performing a 3D fast Fourier transform (FFT) on the 3D reconstruction (SVSEM 3D data stack) of both the DG and sponge regions allows in the former case, determination of the distorted cubic (triclinic) unit cell parameters and in the latter, assessment of the spatial structural correlations within the sponge phase. As described in ref. (20), a Hanning window was applied to the raw SEM 3D data in order to reduce artifacts associated with region of interest edge boundary discontinuities in the subsequent FFT. 2D sections of the 3D FFT were obtained and analyzed via Mathematica notebooks.

**Skeleton network construction.** Tubular minority domains are segmented from the grayscale SVSEM image stack using FIJI (https://imagej.net/software/fiji), with the binarization threshold chosen to match the experimentally reported minority volume fraction (41%). We follow the same procedure used in Ref (20). The binarized volume is reduced to a 1D skeletal curve using FIJI's built-in skeletonization algorithm (63). The resulting voxel skeleton is converted to a graph by removing dangling branches, collapsing voxel chains into single graph edges, and merging junction clusters into single vertices, while preserving the topology of the original tubular network. Node and strut positions are then optimized to lie at positions of maximal PDMS density, corresponding to the central axes of tubular domains and centers of inter-tubular junctions.

**Chirality analysis.** Local chirality is quantified by a dihedral angle $\psi$ measured for consecutive edge triplets in the skeletal graph, as described in (46). For four vertices $(\mathbf{v}_1, \mathbf{v}_2, \mathbf{v}_3, \mathbf{v}_4)$ connected by three consecutive edges with directions $\hat{\mathbf{r}}_1, \hat{\mathbf{r}}_2$ and $\hat{\mathbf{r}}_3$, $\psi$ is computed from the rotation of normals to the planes spanned by adjacent edge pairs along $\hat{\mathbf{r}}_2$. A bond-averaged dihedral angle is assigned to each interior edge $(i, j)$ by averaging $\psi$ over all length-3 paths for which that edge is the middle edge.

**IMDS shape analysis.** Measurements of the IMDS shape are computed based on a generated level set function of the normalized PDMS intensity. The raw SVSEM image stack is binarized (via a threshold to match the PDMS volume fraction), smoothened via a Gaussian filter of 3-4 voxels and then input to a $2^{nd}$-order interpolating function over the sample volume. The 50% isocontour of this smooth interpolation is used to compute a triangulated mesh of the IMDS (with maximum face area of 10 $nm^2$). Normals to the IMDS are computed from the gradient of the intensity level set, and mean and Gaussian curvature are computed from interpolated first and second derivatives of the intensity using standard relations for level set geometry.

**Mesoatom analysis.** As proposed previously, mesoatoms of PCN correspond to volumes occupied by groupings of macromolecules that are clustered at nodal interconnections of tubular network domains (5, 15). As a proxy of for actual volumes occupied by chains, we compute the medial map of the IMDS, which maps each point in the sample volume **p** onto the nearest point on the IMDS, **m**(**p**) (47). The set of points **p** mapping to the same IDMS point **m**(**p**) extend along its normal at **m**(**p**) provide a proxy for entropy maximizing configurations of diblock chains (i.e. straight paths, normal to the IMDS) of both chemical subdomains and pass through the IMDS at their inter-block junction (48). Based on the discretized mesh of the IMDS extracted from SVSEM, we compute the inner (PDMS region) and outer (PS region) medial surfaces. Our approach, which is based on and adapts the algorithm described in ref. (47) for use on voxelized data, uses the Voronoi tessellation generated by IMDS vertices (see **SI appendix** for additional detail).This results in space filling tessellation of triangular, prismatic “wedges”, each extending normally from a triangular facet of the IMDS to triangular facets on both the inner and outer medial surfaces. Thus, each straight wedge includes an inner (PDMS) and outer (PS) block region, and the local thickness of each subdomain corresponds to the appropriate length of that wedge element. Mesoatoms are extracted by finding the set of wedges whose inner edges (facets on the PDMS medial surface) are closest to a given node of the skeletonized PDMS network.

**Loop and linking analysis.** Loops are identified as simple cycles in the skeletal graph and computed for loop lengths up to 12. Linking number between pairs of loops are computed using a Gauss linking integral (78) for discretized polygonal loops, summing over all pairs of line segments between the two polygonal loops. Loop pairs with nonzero linking number are classified as linked. To characterize how linking varies spatially across the sample, each edge $e$ is assigned a linking fraction $f(e) = | C_{\text{linked}}(e) | / | C(e) |$, where $C(e)$ is the set of loops of length $l \leq 12$ containing edge $e$, and $C_{\text{linked}}(e)$ is the subset of loops containing edge $e$ that is linked to at least one other loop.

## Supporting Information for

# Topological and morphological signatures of disorder in block copolymer sponge networks

Xueyang Feng, Suman Kulkarni, Micheal S. Dimitriyev, Dani S. Bassett, Randall D. Kamien, Edwin L. Thomas, Gregory M. Grason

Gregory M. Grason
Email: grason@umass.edu

### 1. Supplemental Methods Appendix

**A. Constructing the network skeleton.** We use the image-analysis software FIJI (https://imagej.net/software/fiji) to identify tubular domains formed by the minority component. Starting from the grayscale 3D image stack, we binarize the volume and choose an intensity threshold so that the resulting binary segmentation matches the experimentally reported minority volume fraction ($40\%$ minority volume fraction), thereby separating minority and majority components. We then use the inbuilt skeletonization feature in FIJI to reduce the binarized 3D volume data into a 1D curve that is a collection of voxels (1). To convert the 1D voxel curve into a skeletal graph, we perform additional post-processing using a custom analysis pipeline in Mathematica. We first take the voxel coordinates as vertices and connect each voxel to its adjacent neighbors in a $3 \times 3 \times 3$ voxel neighborhood. We then identify vertices that lie at the boundary and hold them fixed, and iteratively remove vertices in the interior that have only one nearest neighbor. This eliminates dangling branches of the 1D curve. Finally, we convert the remaining curve into straight bonds by iteratively removing vertices with two neighbors: whenever a vertex has exactly two neighbors, we delete it and connect its two neighbors directly, thereby collapsing long voxel chains into single graph edges while preserving topology. The end of this process typically results in small clusters of nearby vertices at junctions (nodes). We merge such clusters into a single vertex. At the end of this procedure, we obtain a skeletal graph with the same topology of the tubular network that we started out with.

While the resulting skeletal graph approximates the middle of the minority block domains, the decimation of the voxelized skeleton, as described above, results in straight struts that are somewhat offset from the middle of the domains. In order to adjust the skeletal network such that it is a "best fit" to the center of the minority block domain, we perform an additional optimization step, which is described in Prasad, et al. (2). In short, skeletal network vertices are adjusted until the integrated composition $\sum_{\langle ij \rangle} \int_i^j \mathrm{d}s\, \phi(x(s))/L$ is maximized. This optimization is implemented in Matlab using the *fmincon* function, which numerically minimizes constrained nonlinear functionals.

**B. Measuring network chirality.** To quantify chirality of the discrete networks, we measure a dihedral angle for every sequence of three consecutive edges in the network (2). Consider four vertices $(\mathbf{v}_1, \mathbf{v}_2, \mathbf{v}_3, \mathbf{v}_4)$ connected by three consecutive edges, and let the corresponding edge directions be $\mathbf{r}_1 = \mathbf{v}_2 - \mathbf{v}_1$, $\mathbf{r}_2 = \mathbf{v}_3 - \mathbf{v}_2$, and $\mathbf{r}_3 = \mathbf{v}_4 - \mathbf{v}_3$. The normals to the two planes spanned by the adjacent edge pairs are

$$\hat{\mathbf{n}}_{12} = \frac{\hat{\mathbf{r}}_1 \times \hat{\mathbf{r}}_2}{|\hat{\mathbf{r}}_1 \times \hat{\mathbf{r}}_2|} \quad \text{and} \quad \hat{\mathbf{n}}_{23} = \frac{\hat{\mathbf{r}}_2 \times \hat{\mathbf{r}}_3}{|\hat{\mathbf{r}}_2 \times \hat{\mathbf{r}}_3|} \ . \qquad \text{[SE1]}$$

We measure the dihedral angle $\theta_d$ using

$$\sin\theta_d = (\hat{\mathbf{n}}_{12} \times \hat{\mathbf{n}}_{23}) \cdot \hat{\mathbf{r}}_2 \quad \text{and} \quad \cos\theta_d = \hat{\mathbf{n}}_{12} \cdot \hat{\mathbf{n}}_{23} \,. \qquad \text{[SE2]}$$

This dihedral angle provides a local sense of rotation for the three consecutive edges. We then define a *bond-averaged dihedral angle* for each (non-boundary) edge by averaging $\theta_d$ over all length-3 paths for which that edge is the middle edge. That is, for a given edge $(i,j)$, we enumerate all valid choices of a neighbor of $i$ (excluding $j$) and a neighbor of $j$ (excluding $i$), forming a triplet of consecutive edges that pass through edge $(i,j)$. We compute $\theta_d$ for each such triplet and then average these values to assign a single bond-averaged dihedral angle to edge $(i,j)$.

**C. Loop and linking analysis.** To characterize the topology of the skeletal networks, we examine their looping structure. A loop is defined as a simple cycle: a closed walk that returns to its starting vertex without repeating any other vertex. In the ordered region, the skeleton network consists of two separate inter-threading networks of srs topology. The shortest loop passing through any vertex in the ordered srs graph has length 10, consistent with descriptions of the Laves/srs graph (3). In contrast, in the amorphous region loops are variable in length. We enumerate all loops of length up to 12 in the skeletal graph in Mathematica using the in-built function *FindCycle*.

As described in ref. (4) the linking number L between two closed curves $\Gamma$ and $\Gamma'$, parameterized by $\boldsymbol{R}(s)$ and $\boldsymbol{R}'(s')$ (with length $L$ and $L'$, is given by the Gauss linking integral

$$L(\Gamma,\Gamma') = \frac{1}{4\pi}\oint_0^L \mathrm{d}s \oint_0^{L'} \mathrm{d}s' \left[\frac{\mathrm{d}\boldsymbol{R}}{\mathrm{d}s} \times \frac{\mathrm{d}\boldsymbol{R}'}{\mathrm{d}s'}\right] \cdot \frac{[\boldsymbol{R}(s) - \boldsymbol{R}'(s')]}{||\boldsymbol{R}(s) - \boldsymbol{R}'(s')||^3}\,. \qquad \text{[SE3]}$$

In practice, each closed loop is a polygon with edges along the skeletal network, and we compute Eq. [SE3] by summing the Gauss integral over all pairs of line segments in the two polygons — one segment from loop $\Gamma$ and one segment from loop $\Gamma'$. We use this procedure to compute linking numbers for loop pairs in the skeletal network, and we classify a pair of loops as linked when the resulting linking number is nonzero.

To examine how the linking structure varies across the sample, we assign each edge a *linking fraction* defined as the fraction of loops that involve the edge which are linked to at least one other loop. To be more explicit: For each edge $e$, let $\mathcal{C}(e)$ denote the set of calculated loops (here, loops of length $l \leq 12$) that contain edge $e$, and $\mathcal{C}_{\text{linked}}(e)$ denote the subset of these loops that are linked with at least one other loop. For all edges with at least one loop passing through, the edge linking fraction is

$$f(e) = \frac{|\mathcal{C}_{\text{linked}}(e)|}{|\mathcal{C}(e)|}. \qquad \text{[SE4]}$$

**D. Extracting terminal boundaries from medial axis transform.** A block copolymer is taken to be "associated" to a given patch of intermaterial dividing surface (IMDS) if it passes through the IMDS, with the A-B junction typically positioned on or close to the IMDS (5). The domain of an IMDS patch is then taken to be the region of space where chains are most likely to be associated with a given IDMS patch if selected at random. The domain boundaries, or *terminal boundaries* $\mathcal{T}$, correspond to the transition of chain association, where a chain is equally likely to be associated to two separate IMDS patches.

Given that the definition of terminal boundaries requires an evaluation of chain association probabilities, there is considerable difficulty in both their theoretical and experimental determination. However, it has been found that the *medial surfaces* generated by an IMDS-like surface serve as reasonable approximations to the terminal boundaries (6–8).

Similar to a Voronoi tessellation, the medial axis transformation identifies regions of space that are closest to a given "generating set" according to the Euclidean distance. Unlike the Voronoi tessellation, which is defined for a discrete set of generating points, the medial axis transformation generalizes generating sets to continuous surfaces $\mathcal{G}$. Let $\widehat{\mathbf{N}}$ be the surface normal; this is defined up to a choice of sign. The generating surface $\mathcal{G}$ then, under the medial axis transform, maps to two medial surfaces $\mathcal{T}_-$ and $\mathcal{T}_+$: one for each sign of $\hat{\mathbf{N}}$. These medial surfaces are taken to be approximations of the terminal boundaries. Note that while the generating surface $\mathcal{G}$ resembles the IMDS, slight variations are required to ensure local volume balance.

The algorithm used here, based on the method discussed in ref. (9), relies on the creation of a triangular mesh discretization of $\mathcal{G}$. The mesh discretization consists of a set of vertices $p_i \in \mathcal{G}$ and faces $\{f_{ijk}\}$, where each triangular face is a triplet of vertex indices $f_{ijk} = \{i, j, k\}$, creating triangle $\triangle\, p_i p_j p_k$. A Voronoi tessellation $\mathcal{V}$ is then created using these vertices, consisting of a collection of cells $\mathcal{V} = \cup_i \mathcal{V}_i$ with each cell $\mathcal{V}_i$ containing the set of points in space closest to vertex $p_i$.

In the usual medial axis transformation algorithm, a line $\boldsymbol{\ell}_i$ is created passing through vertex $p_i$, oriented along normal direction $\hat{\mathbf{N}}_i = \hat{\mathbf{N}}(p_i)$. This line then intersects with Voronoi cell $\mathcal{V}_i$ at two points: $p_i^-$ along the $-\hat{\mathbf{N}}_i$ direction and $p_i^+$ along the $+\hat{\mathbf{N}}_i$ direction. These points lie on either medial surface, $p_i^\pm \in \mathcal{T}_\pm$. However, this approach is ill-suited for voxelated experimental data, since the locations of the intersection points are highly susceptible to small errors in the direction of the surface normal $\hat{\mathbf{N}}_i$. Instead, we consider an *approximate medial axis transformation* in which $p_i^\pm$ are chosen to be the vertices of the Voronoi cell $\mathcal{V}_i$ that are maximally distant from the generating vertex $p_i$ on either side of the generating surface. The surfaces obtained from this approximation will approach the true medial surfaces as the density of generating points increases on a smooth generating surface; in this limit, the Voronoi cells become increasingly narrow with the length aligned along the normal of the generating surface. Since the approximate method only uses the surface normal to distinguish between the medial surface vertices $p_i^\pm$, it is robust to non-smooth gradients in $\hat{\mathbf{N}}_i$ arising from measurement resolution. Since the medial surface vertices $p_i^\pm$ have the same neighbors as the generating surface vertices $p_i$, the indices in faces $f_{ijk}$ can be used to construct pairs of faces on the two medial surfaces, thus defining suitable triangle mesh discretizations of the medial surfaces $\mathcal{T}_\pm$.

## Figures

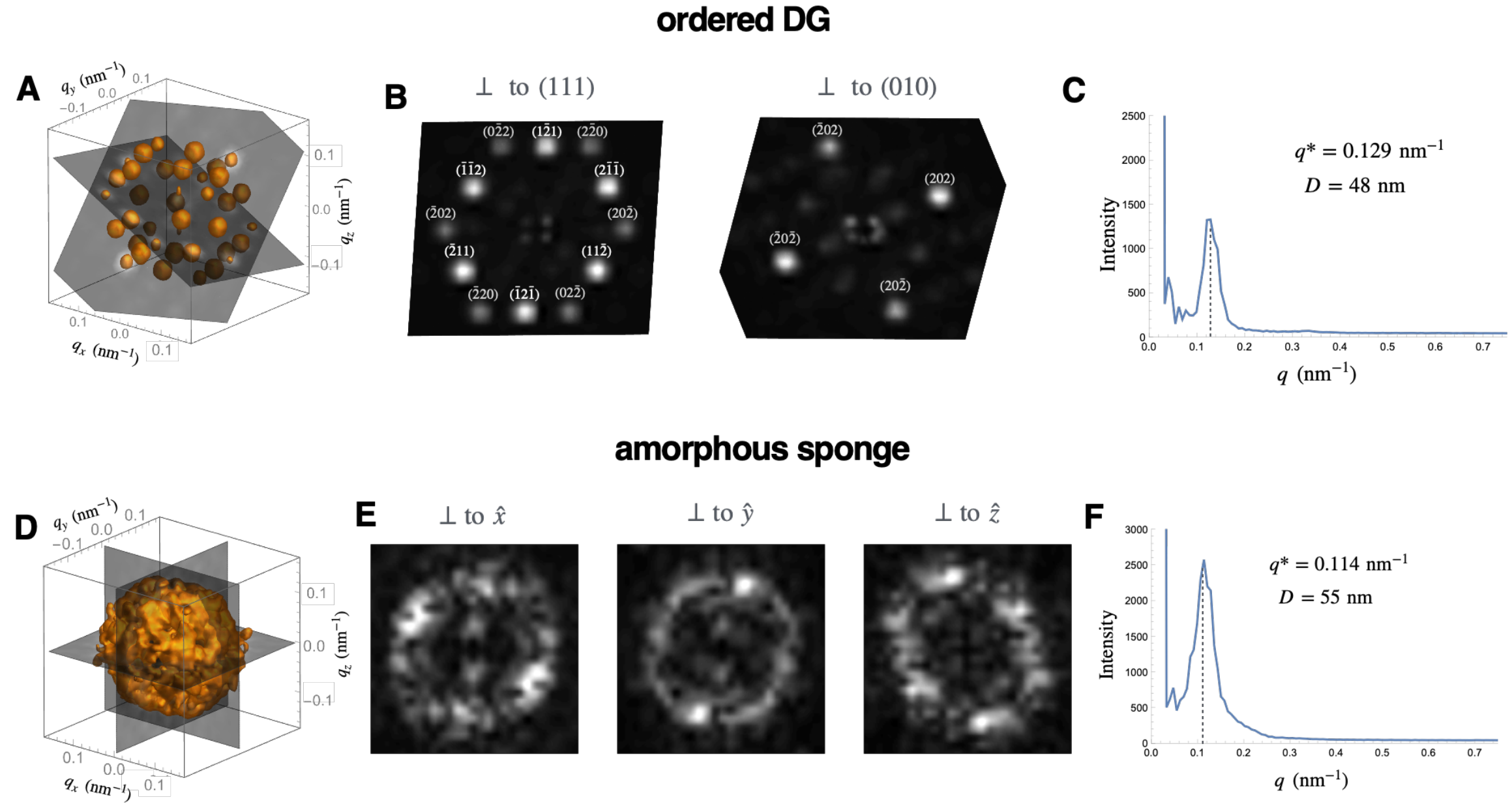

**Fig. S1. Fourier analysis (2D sections and radial intensity profiles)**. (A) and (D) show isocontour plots of 3D FFT of SVSEM data of ordered DG and amorphous sponge regions, respectively. Additionally 2D slices through 3D FFT are shown corresponding to the sections shown in (B) and (E), where intensity values are plotted on a linear scale, and in (B) bright intensity spots are indexed according to reflections of a (triclinically distorted) $\boldsymbol{Ia\bar{3}d}$ structure. (C) and (F) show profiles of orientationally averaged Fourier intensity versus wavevector for DG and sponge, respectively, with vertical dashed line highlighting the location of the peak intensity.

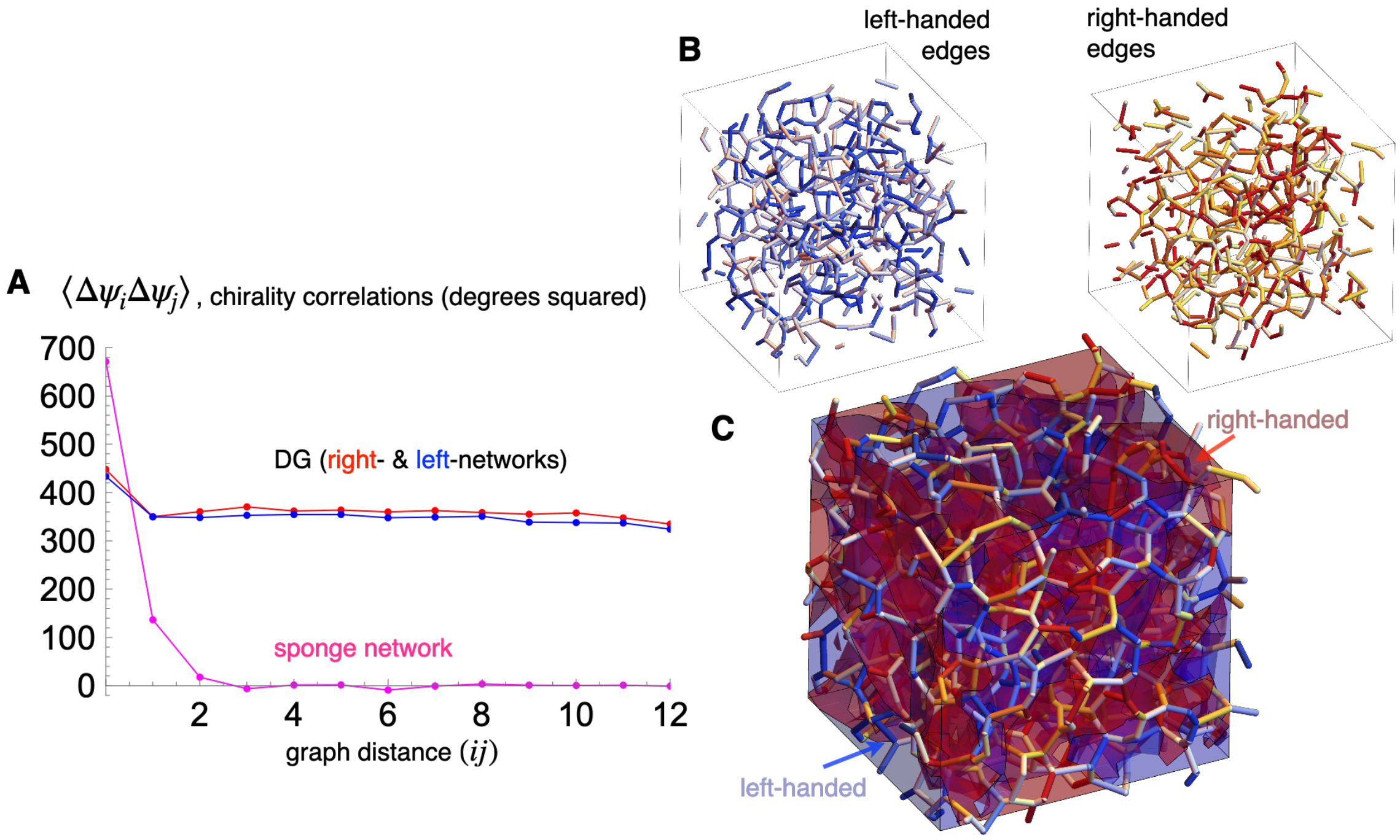

**Fig. S2. Intra-network (dihedral) chirality correlations.** (A) Plot of correlation functions of edge chirality versus intra-network graph distance between edges, for both PDMS networks of DG and the single PDMS network of amorphous sponge. Chirality at strut *i* is measured by the excess dihedral twist $\Delta\boldsymbol{\psi}_i = \mathbf{90}^{\circ} - \boldsymbol{\psi}_i$ . DG shows long-range correlations in each network, while sponge exhibits local excess of like-chirality within a 2-3 neighbor edges. (B) shows the respective left-handed (blue) and right-handed (edges) of the amorphous sponge. The spatial structure of chiral correlations in sponge is highlighted as transparent blue or red, showing alternation between locally right- and left-handed dihedral twist on the scale of 2-3 edges.

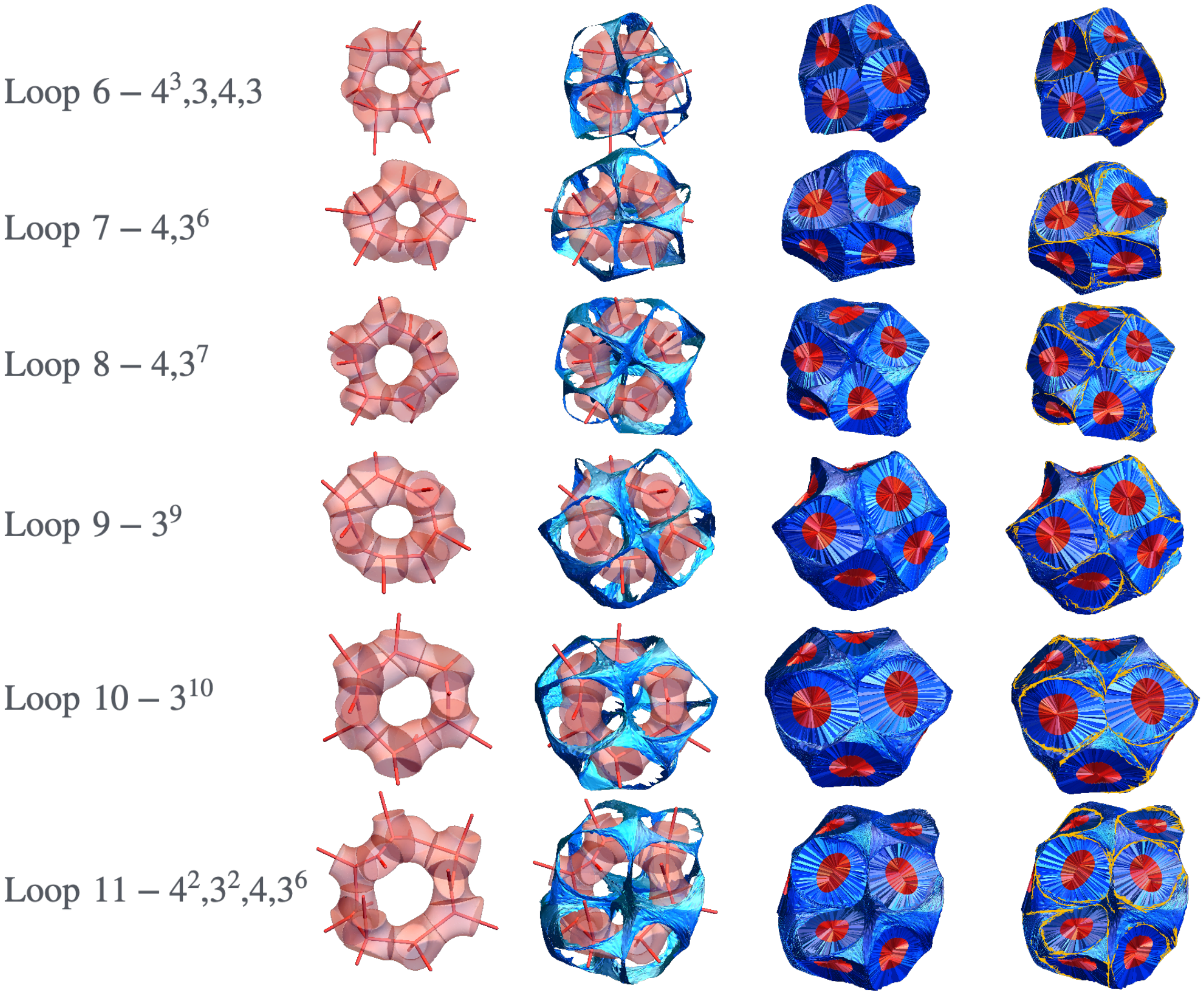

**Fig. S3. Loop examples of random sponge.** Graphical table of loops of variable size from loops of 6 to 11 mesoatoms, rendered highlight PDMS domain and skeleton (red), PS domain boundary (blue) and volume-filling mesoatoms, with sharply-folded, cusps highlighted in yellow. Loop composition (i.e. number of 3- vs. 4-valent nodes) is given in the notation of ref. (10) .

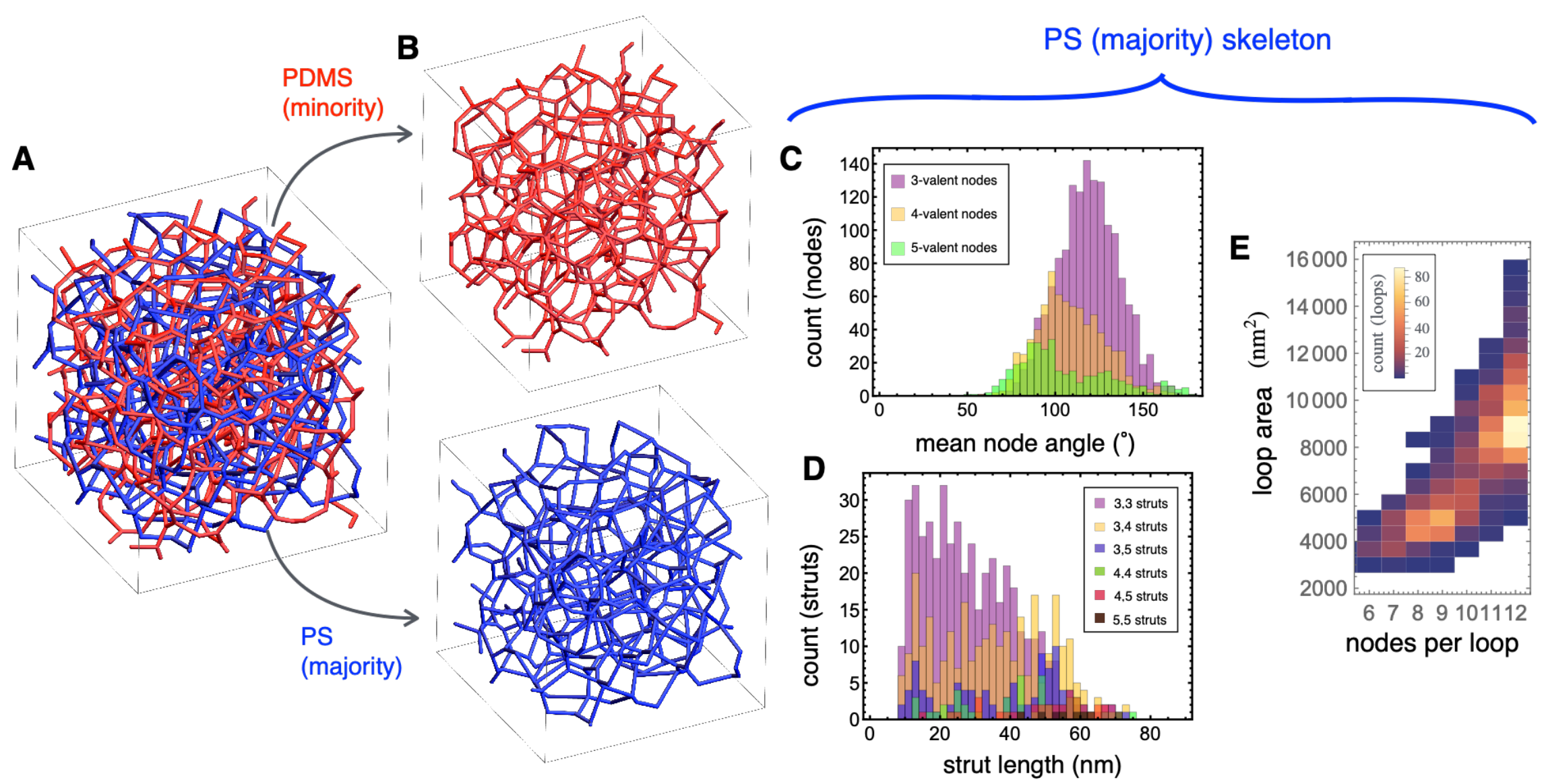

**Fig. S4. Majority component (PS) network random sponge.** (A) shows intercalated skeletons of PDMS (red) and PS (blue) domain component of random sponge phase, with separated networks in (B). (C), (D) and (E) show respective histograms for mean internal node angle, strut length and loop areas for PS network of sponge. Node composition of PS network is 75% 3-valent, 19% 4-valent and 6% 5-valent.

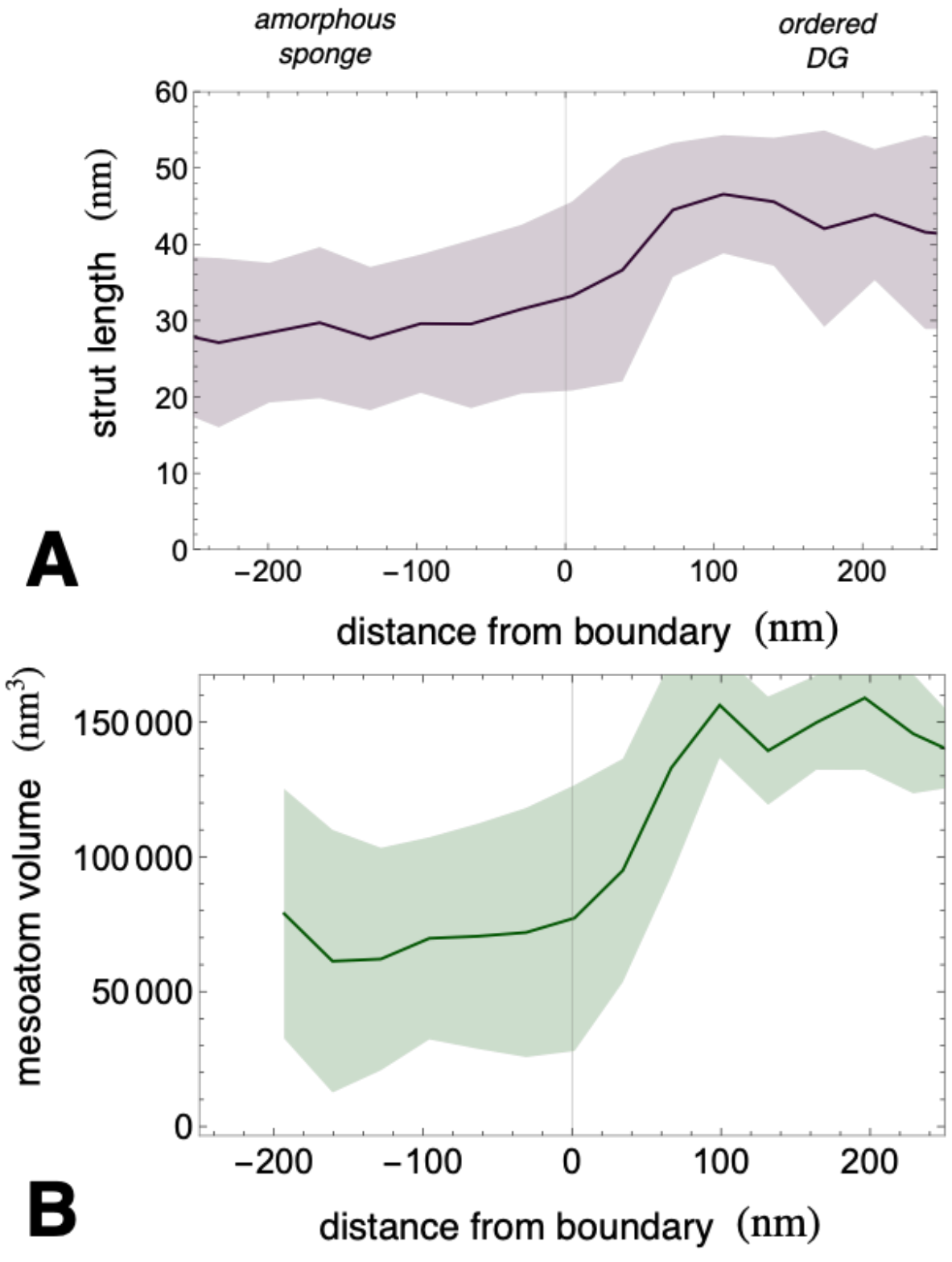

**Fig. S5. Transition of network and mesoatom geometry across boundary**. In (A), plot of mean length of PDMS network struts for variable distance from the ordered DG and random sponge boundary, defined as described in main text. In (B), plot of mean mesoatom volume for variable distance from order/disorder boundary. In both, the standard deviation of the data range is shown as semi-transparent color.

**Movie S1 (Video1_OrderedvsAmorphous.mp4).** Rotating views of SVSEM rendering of ordered DG (left) and amorphous sponge (right) morphologies corresponding to cubic regions of 300 nm on a side. Initially, PDMS and PS regions are colored red and blue, respectively, followed by an overlaid skeleton of PDMS network, which is colored according to the mean dihedral rotation angle.

**Movie S2 (Video2_mesoatoms.mp4).** Rotating views of example mesoatomic region extracted from SVSEM reconstructions based on the medial transformation. PDMS and PS regions are colored red and blue respectively. PDMS surface regions correspond to cuts through the tubular struts of the tubular minority network, and hence, only appear on surfaces dividing neighboring mesoatoms in the same network. All mesoatoms of ordered DG are trivalent, but correspond to left- or right-handed dihedral network chirality, whereas mesoatoms of the amorphous sponge are mostly (91%) trivalent and include a small fraction (9%) of tetravalent units.

**Movie S3 (Video3_loops_mesoatoms.mp4).** Rotating views of example loops, first showing volume filling mesoatoms and then successively removing the PS (blue) and PDMS (red) wedge-like volumes of medial tessellation to reveal the corresponding regions occupied by the tubular PMDS subdomain (transparent red) and underlying skeletal networks (red). On the left, two interlinked loops of 10 from the ordered DG are shown. On the right, a 9-mesoatom loop from the amorphous sponge, showing the unlinked character of the disordered single-network morphology.

**Movie S4 (Video4_interface_mesoatomwall.mp4).** An animated view of the mesoatomic regions at the boundary between the ordered DG and amorphous sponge region. Initially, the boundary mesoatoms are superposed on a view of the PDMS skeletons colored according to linking fraction (as in Fig. 5). The boundary view is then rotated to show face-on views from the ordered region and then from the amorphous sponge region. These views highlight differences in the smooth (DG) versus cuspy (sponge) features of the PS boundary surfaces in mesoatomic shapes on either side of this narrow (~1-2 mesoatom thick) region).

**SI References**